\documentclass[11pt,a4paper]{article}
\usepackage{cite}
\usepackage{amsmath, amsthm, amssymb,slashed}
\usepackage{ifpdf}
\ifpdf
  \usepackage[pdftex]{graphicx}
  \usepackage{epstopdf}
\else
  \usepackage[dvips]{graphicx}
\fi
\parskip=\baselineskip
\def\Bbb{\mathbb}
\def\Tr{{\rm Tr}}
\def\16{{\bf 16}}
\def\1{{\bf 1}}
\def\2{{\bf 2}}
\def\3{{\bf 3}}
\def\4{{\bf 4}}
\def\be{\begin{equation}}
\def\ee{\end{equation}}

\def\bar{\overline}

\def\R{{\Bbb{R}}}\def\Z{{\Bbb{Z}}}

\def\hat{\widehat}
\font\teneurm=eurm10 \font\seveneurm=eurm7 \font\fiveeurm=eurm5
\newfam\eurmfam
\textfont\eurmfam=\teneurm \scriptfont\eurmfam=\seveneurm
\scriptscriptfont\eurmfam=\fiveeurm

\font\teneusm=eusm10 \font\seveneusm=eusm7 \font\fiveeusm=eusm5
\newfam\eusmfam
\textfont\eusmfam=\teneusm \scriptfont\eusmfam=\seveneusm
\scriptscriptfont\eusmfam=\fiveeusm

\font\tencmmib=cmmib10 \skewchar\tencmmib='177
\font\sevencmmib=cmmib7 \skewchar\sevencmmib='177
\font\fivecmmib=cmmib5 \skewchar\fivecmmib='177
\newfam\cmmibfam
\textfont\cmmibfam=\tencmmib \scriptfont\cmmibfam=\sevencmmib
\scriptscriptfont\cmmibfam=\fivecmmib

\numberwithin{equation}{section}

\def\d{\mathrm d}
\def\C{{\Bbb C}}
\def\Z{{\Bbb Z}}

\def\bar{\overline}

\begin{document}
\begin{titlepage}
\begin{flushright}

\end{flushright}
\vskip 1.5in
\begin{center}
{\bf\Large{Integrable Lattice Models From Gauge Theory}}
\vskip
0.5cm {Edward Witten} \vskip 0.05in {\small{ \textit{School of
Natural Sciences, Institute for Advanced Study}\vskip -.4cm
{\textit{Einstein Drive, Princeton, NJ 08540 USA}}}
}
\end{center}
\vskip 0.5in
\baselineskip 16pt
\begin{abstract}
These notes provide an introduction to recent work by Kevin Costello in which integrable lattice models of
classical statistical mechanics
in two dimensions are understood in terms of quantum gauge theory in four dimensions.  This construction will be compared to the more
familiar relationship between quantum knot invariants in three dimensions and Chern-Simons gauge theory.   (Based on a 
Whittaker Colloquium at the University of Edinburgh and a lecture at Strings 2016 in Beijing.)
\end{abstract}
\date{October, 2016}
\end{titlepage}
\def\Hom{\mathrm{Hom}}

\def\Tr{{\mathrm{Tr}}}
\def\CS{{\mathrm{CS}}}
\def\d{{\mathrm d}}
\def\i{{\mathrm i}}
\def\LG{{\mathcal L}G}
\def\R{{\Bbb R}}
\def\C{{\Bbb C}}
\def\Z{{\Bbb Z}}
\def\bar{\overline}

\section{Preliminary Remarks}

Since the discovery of the Bethe ansatz in the early days of quantum mechanics, integrable quantum
spin systems in $1+1$ dimensions and their close cousins of various sorts have been a topic
of much fascination.
Important advances have been made by many physicists and mathematicians, among them
 Onsager, Yang, Baxter, Lieb, Kruskal, Fadde'ev,
Drinfeld,  Miwa, Jimbo,
and A. and Al. Zamolodchikov.  The subject is so multi-faceted that no short summary can do justice to it.

In today's lecture, I will be describing a new approach to integrable lattice models of two-dimensional
classical statistical mechanics, developed recently by Kevin Costello \cite{Costello}.   Costello's work has offered an essentially new perspective in which these models
are understood in terms of four-dimensional gauge theory.\footnote{Another and superficially quite different relation of some of the same models to gauge theory
was discovered earlier by Nekrasov and Shatashvili \cite{NS,NS2}.}  Arguably this perspective is in line with a vision
relating theories in different dimensions that was offered many years ago by Michael Atiyah \cite{Atiyah}.  As we will
see, Costello's work is also a close cousin of the relationship \cite{Witten} between Chern-Simons gauge theory in
three dimensions and the Jones polynomial of a knot.\footnote{Costello has actually described two related 
approaches to this
subject, one in terms of a four-dimensional cousin of Chern-Simons gauge theory and one in terms of a twisted version
of four-dimensional $\mathcal N=1$ super Yang-Mills theory.  Roughly speaking, the first approach, which we follow here,
comes by integrating out some variables from the second.  For a review of the
second approach, see \cite{Costello2}.}   

There are several different kinds of integrable systems in one space or two spacetime dimensions,
including classical nonlinear PDE's, continuum quantum field theories, quantum spin chains, and classical
lattice systems.  However, these different types of model turn out to be closely related.
(Among 
the many excellent sources where one can find different perspectives are \cite{Faddeev, Jimbo,SpecialCase,PY}.)
Even though our main topic today will be lattice statistical mechanics in two dimensions, to motivate
the ideas I will begin with integrable models of
continuum quantum field theory in $1+1$ dimensions.   The ideas that I will sketch were first used as a tool to directly construct
a relativistic $S$-matrix by A. and Al. Zamolodchikov \cite{Zam}.

\begin{figure}
 \begin{center}
   \includegraphics[width=2.5in]{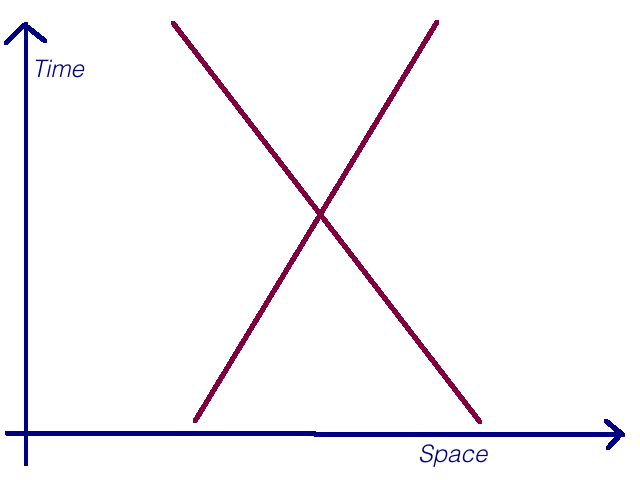}
 \end{center}
\caption{\small A spacetime picture of two-body elastic scattering in $1+1$ dimensions.}
 \label{spacetime2}
\end{figure}

Fig. \ref{spacetime2} is meant to be a spacetime picture of elastic scattering of two particles in $1+1$ dimensions.  A particle
of constant velocity is represented by a straight line, with a slope depending on the velocity.
   Because of conservation of energy and momentum, the outgoing particles go off at the same slope (same
   velocity) as the incoming particles.  There are time delays that I have not tried to draw.  The time delays mean that
   the outgoing lines are parallel to the ingoing ones, but displaced slightly inwards.  This will not really affect our
   discussion.

\begin{figure}
 \begin{center}
   \includegraphics[width=3in]{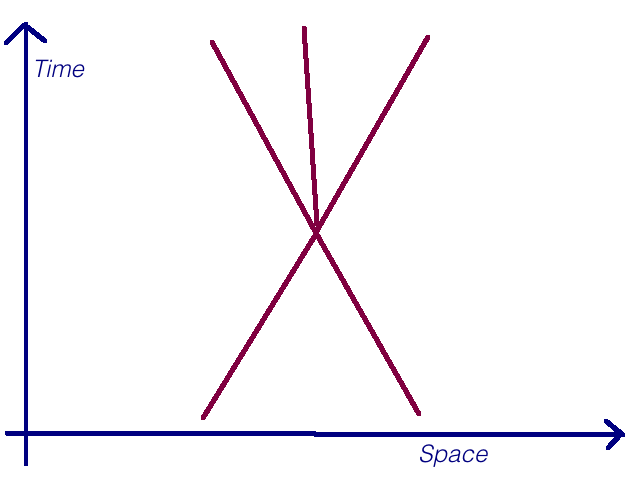}
 \end{center}
\caption{\small Inelastic scattering in $1+1$ dimensions with particle production.}
\label{SpaceTime3}
\end{figure}

In a typical relativistic quantum field theory, there are also particle production processes, which are a large part of what makes  quantum field theory interesting.   An example with two particles going to three is sketched in fig. \ref{SpaceTime3}.
     The symmetries of typical relativistic field theories allow such processes and they happen all the time in 
     the real world.    In a $2\to 3$ scattering event, in a massive theory, the incoming and outgoing lines can be assumed
     to all end or begin at a common point in spacetime, to within an error that depends on the range of the particle
     interactions and is reflected in the time delays.
     
\begin{figure}
 \begin{center}
   \includegraphics[width=3in]{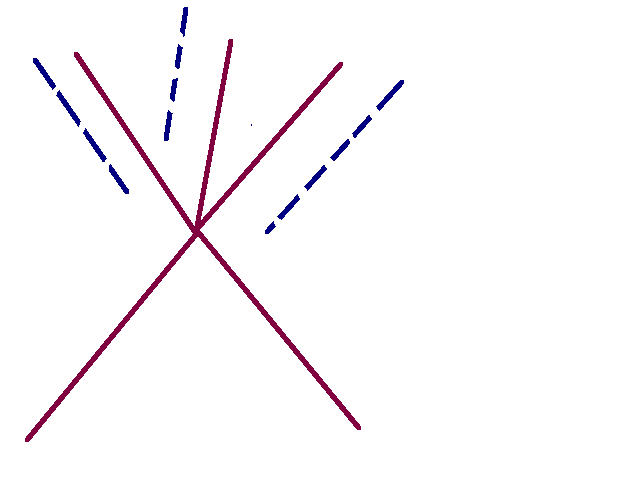}
 \end{center}
\caption{\small An inelastic $2\to 3$ scattering process is sketched by the solid lines.  In a theory that has
a symmetry that moves particles by an amount that depends on their momentum, the application of such a symmetry
would move the outgoing lines parallel to themselves, leaving their slope unchanged. (By combining the symmetry in
question with an ordinary spacetime translation, we can assume that  the incoming lines are unaffected, as drawn here.) After such a transformation,
the outgoing lines -- now showed as dotted lines -- have no common point of intersection in the plane, let along a common
point of intersection at which they also meet the incoming lines.  This means
that a $2\to 3$ process is not possible in a theory with such a symmetry.}
 \label{Prod2}
\end{figure}

However, in two spacetime dimensions, there are ``integrable'' field theories that have extra symmetries that  commute
with the velocity or momentum but move a particle in space by an amount that
depends on its velocity.   Then particle production is not possible.  Starting with a  spacetime history in
which the incoming and outgoing lines meet at a common point in spacetime, a symmetry that moves  the incoming and outgoing
lines by a velocity-dependent amount will create a history such as that of fig. \ref{Prod2} in which the outgoing particles could
have had no common origin in spacetime. 


\begin{figure}
 \begin{center}
   \includegraphics[width=2.5in]{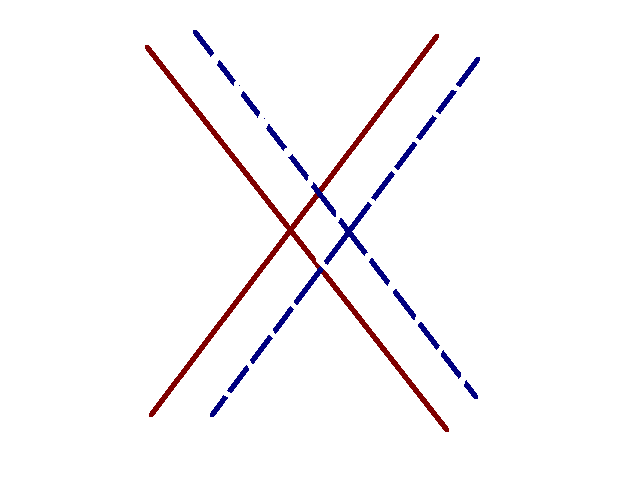}
 \end{center}
\caption{\small Two lines in the plane generically intersect, a statement that is not affected by translating them parallel to themselves.
So two-body elastic scattering can be nontrivial even in an integrable field theory.}
\label{Twop2}
\end{figure}

By contrast, two particle  scattering does happen even in integrable systems, since two lines in the plane do generically intersect, as in fig.
\ref{Twop2}.   But in an integrable theory, two particle scattering is purely {\it elastic}, in the sense that the initial and final particles
have the same masses (this is true separately for the left- and right-moving particles).  Otherwise, the initial and final velocities are different  and consideration of a symmetry
that moves particles in a velocity-dependent way again leads to a contradiction.

   \begin{figure}
 \begin{center}
   \includegraphics[width=2.5in]{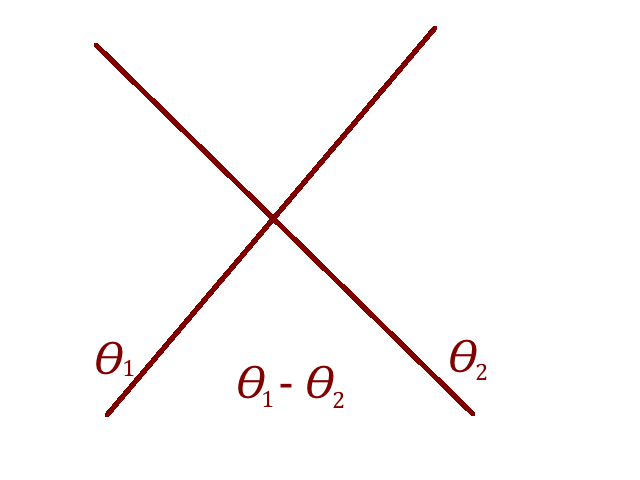}
 \end{center}
\caption{\small Elastic scattering depends only on the rapidity difference.}
\label{Rapidity}
\end{figure}

How do we characterize a particle?    A particle has a velocity, or better, in relativistic terms, a ``rapidity''  $\theta$.   The energy $E$
and momentum $p$ are expressed in terms of $\theta$ and the particle mass $m$ by 
\be \begin{pmatrix} E\cr p\end{pmatrix}= m \begin{pmatrix} \cosh\theta \cr \sinh\theta\end{pmatrix}\ee
 A Lorentz boost adds a constant to the rapidity, so scattering of two particles with rapidities $\theta_1$ and 
 $\theta_2$ depends only on the rapidity difference $\theta=\theta_1-\theta_2$ (fig. \ref{Rapidity}).

But the amplitude for scattering of two particles of rapidities $\theta_1$ and $\theta_2$ is in general 
not only a function of the rapidity difference $\theta=\theta_1-\theta_2$  because there may be several 
different ``types'' of particles of the same mass. 
An obvious reason for this, although not the most general possibility, is that the theory might have a symmetry group $G$ and the particles may 
be in an irreducible representation $\rho$ of $G$.

   \begin{figure}
 \begin{center}
   \includegraphics[width=2.5in]{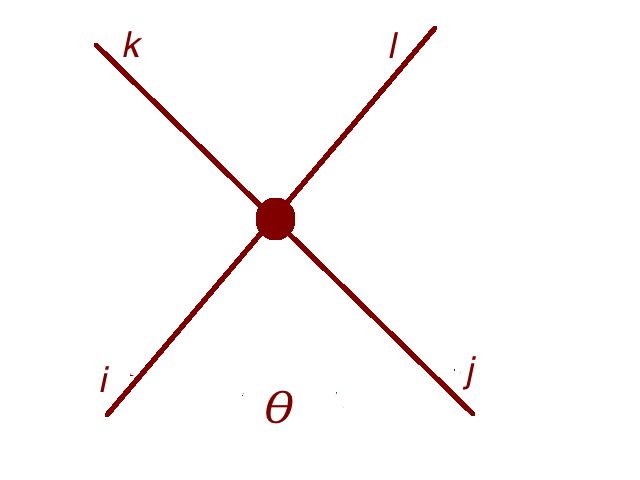}
 \end{center}
\caption{\small Elastic scattering of particles that carry labels $i,j,k,l$, with rapidity difference  $\theta$.}
\label{Rapidity2}
\end{figure}  

The picture is then more like fig. \ref{Rapidity2}.
   Here $i,j,k,l$ can be understood to represent basis
vectors in the representation $\rho$.
We write $R_{ij,kl}(\theta)$ for the quantum mechanical ``amplitude'' that describes this process.   It is usually called the
$R$-matrix.

   \begin{figure}
 \begin{center}
   \includegraphics[width=3in]{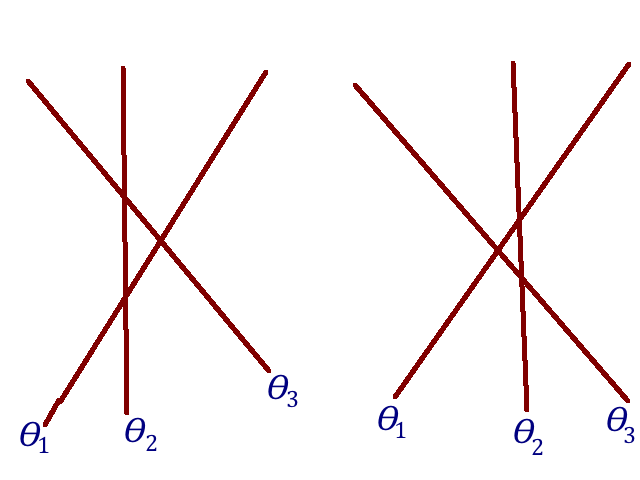}
 \end{center}
\caption{\small There are two ways to express a $3\to 3$ scattering process as a succession of three $2\to 2$ scattering processes.}
\label{YB0}
\end{figure}

   \begin{figure}
 \begin{center}
   \includegraphics[width=3.8in]{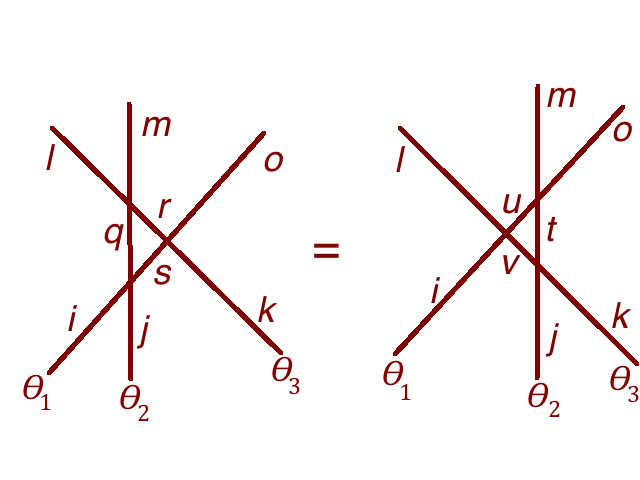}
 \end{center}
\caption{\small In more detail, the equivalence of the two pictures in fig. \ref{YB0} leads to this relation,
called the Yang-Baxter equation.  A sum over internal particle types ($q,r,$ and $s$ on the left and $u,t$, and $v$ on the right) is 
understood.}
\label{Yang-Baxter}
\end{figure}

The real fun comes when we consider three particles in the initial and final state.   Since
we can move them relative to each other, leaving their slopes (or rapidities) fixed, we can assume that there
are only pairwise collisions.   But there are two ways to do this, as in fig. \ref{YB0}, and they must give equivalent results.
Concretely equivalence of these pictures leads to the celebrated ``Yang-Baxter equation,''  
which schematically reads \begin{equation}\label{therf}R_{23}R_{13}R_{12}=R_{12}R_{13}R_{23}.\end{equation}
The Yang-Baxter equation is a good example of a relationship that is much more transparent in terms of a picture (fig. \ref{Yang-Baxter})
than by writing out an algebraic formula in detail.

Actually, there is a subtle but important difference between the $R$-matrix that solves the Yang-Baxter equation and the $S$-matrix
that describes particle scattering in an integrable relativistic field theory.  The reason for this is that the Yang-Baxter equation is
not sensitive to an overall $c$-number factor, that is it is invariant under the transformation $R_{ijkl}(\theta)\to F(\theta) R_{ijkl}(\theta)$,
for any scalar function $F(\theta)$.  (This invariance is manifest in fig. \ref{Yang-Baxter}; on both the left and the right, one has a succession of three scattering events
with rapidity differences $\theta_i-\theta_j$, $1\leq i<j\leq 3$.  So overall factors $F(\theta_i-\theta_j)$ will cancel out.)
In applications of the Yang-Baxter equation to classical statistical mechanics,
 such an overall factor is of little importance and one usually
simply picks a convenient normalization of the solution of the Yang-Baxter equation.   
But in $S$-matrix theory, the prefactor $F(\theta)$ is very important.  In simple examples, as pioneered in \cite{Zam},
the prefactor is determined by conditions of unitarity and crossing symmetry and some knowledge of whether the $S$-matrix should
have bound state poles.   It is possible to have two different models (for example the nonlinear $\sigma$-model with target a sphere $S^{N-1}$ 
and corresponding $O(N)$ symmetry and an $N$-component Gross-Neveu model with the same symmetry) that are governed by the
same $R$-matrix but with different choices of the prefactor.  In fact, recently it has been argued \cite{Zam2} that integrable relativistic field
theories in $1+1$ dimensions have continuous
deformations whose $S$-matrices differ only by the choice of $F(\theta)$.  These deformations depend
on infinitely many parameters that are irrelevant in the renormalization group sense.  

The traditional solutions of the Yang-Baxter equation -- as discovered by Bethe, Lieb, Yang, Baxter, Fadde'ev, Belavin, Drinfeld and others -- 
are classified
by the choice of a simple Lie group $G$ and a representation $\rho$, subject to (1) some restrictions on $\rho$, and (2) the curious fact that in many important cases (like the 6-vertex model of
Lieb and the 8-vertex model of Baxter) a solution of the Yang-Baxter equation associated to a given group $G$ does not actually have $G$ symmetry.    In fact, there are three broad classes of
solutions of Yang-Baxter, which are called rational, trigonometric, and elliptic depending on whether the $R$-matrix is a rational,
trigonometic, or elliptic function of $\theta$.  Only the rational solutions of Yang-Baxter have $G$ symmetry.   Prior to the work
of Costello, it was, at least to me, a longstanding puzzle to understand better why there are solutions of the Yang-Baxter equation that are in some
sense associated to the Lie group $G$ but do not have $G$ symmetry.   (The question can be restated in terms of quantum groups, but to me
this merely raises the equivalent question of ``why'' these exist appropriate quantum group deformations.)

   \begin{figure}
 \begin{center}
   \includegraphics[width=2.5in]{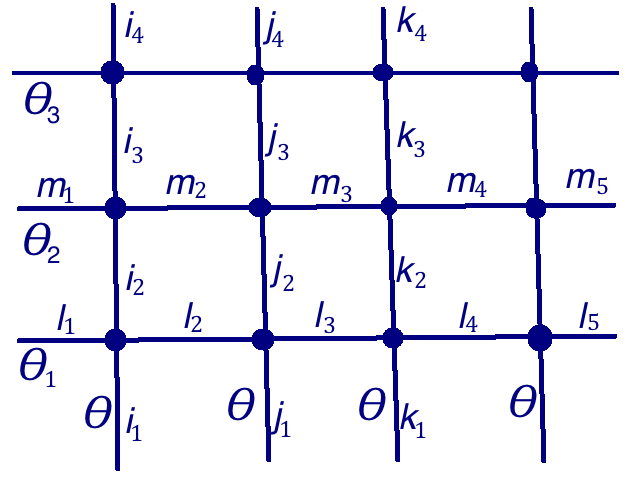}
 \end{center}
\caption{\small An integrable lattice model associated to a solution  of the Yang-Baxter equation.  Horizontal and vertical lines
are labeled by rapidities.  (In constructing the standard integrable models and proving their integrability, one often labels
vertical lines by a common rapidity $\theta$ and horizontal lines by varying rapidities $\theta_i$, so this case is sketched here.)
Line segments between two vertices are labeled by basis vectors $i_s,j_s,k_s$, etc., in some solution of the Yang-Baxter equation; these are
the ``spins'' of the integrable lattice model.  At each
vertex, there is a four-spin interaction given by the relevant matrix element of the $R$-matrix.  In some special cases, the four-spin interactions
collapse to two-spin interactions, leading to more simple-looking integrable models such as the Ising model.
}
\label{Model}
\end{figure}

Now finally we come to the lattice models.
I have 
motivated the Yang-Baxter equation by talking about relativistic scattering, but  solutions of Yang-Baxter can be used  to
construct several
different kinds of integrable model.  Today we will focus on the integrable lattice systems of classical statistical mechanics.  They
are constructed directly from a solution of the Yang-Baxter equation.\footnote{For example, see \cite{SpecialCase}.
The solution of Yang-Baxter used
in this construction  need not necessarily be rational, so for this
application one considers more general solutions of Yang-Baxter than the ones that arise in the integrable relativistic field theories.}
 To understand how this is done, consider fig. \ref{Model}. In this this rather busy picture, the vertical and horizontal lines are labeled by rapidities.  I have labeled the vertical lines by the same rapidity $\theta$ (though this restriction is not necessary) and the horizontal
 lines by different rapidities $\theta_i$.   In addition, each line segment in the figure
 is labeled by a basis vector $i_s,j_s,k_s,\dots$ in some matrix-valued solution of the Yang-Baxter equation.  These labels will be the
 ``spins'' of our  lattice model.  For the Boltzmann weights, we include for each crossing in the figure a factor of
 the appropriate $R$-matrix element.   Thus this is a model in which the spins live on the edges of a square lattice, and there
 is an interaction among every four spins that live on edges that meet at a common vertex.\footnote{To make a physically sensible
 model of classical statistical mechanics, one chooses the rapidities so that the relevant matrix elements of the $R$-matrix are all
 real and positive.}  It turns out that models constructed in this way
 are integrable, because ``the transfer matrices commute,'' which means that (using the
   Yang-Baxter equation) the horizontal lines can be moved up and down past each other. Conversely, familiar integrable lattice
   models can be put in this form.
   
 \section{The Yang-Baxter Equation And Gauge Theory}

   \begin{figure}
 \begin{center}
   \includegraphics[width=2.5in]{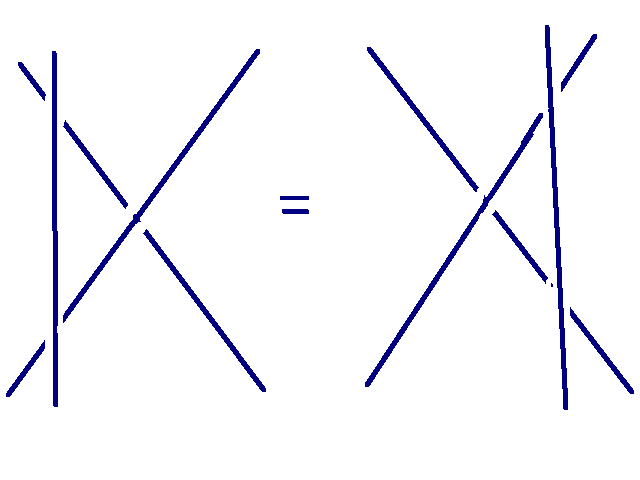}
 \end{center}
\caption{\small A Reidemeister move of knot theory, with an obvious resemblance to the Yang-Baxter equation.}
\label{Reidemeister}
\end{figure}

  \begin{figure}
 \begin{center}
   \includegraphics[width=2in]{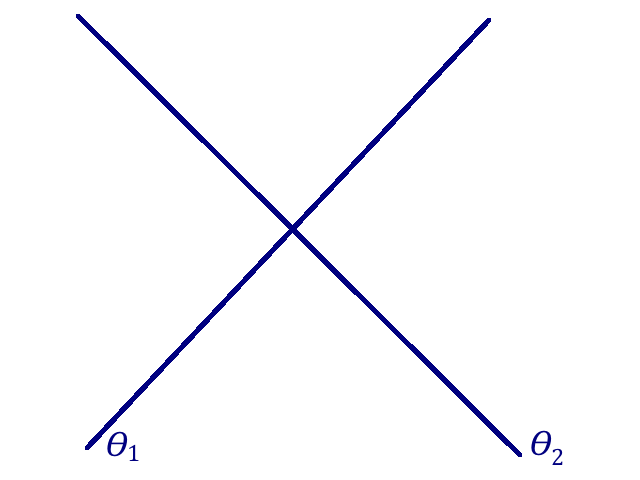}
 \end{center}
\caption{\small In knot theory, one knot passes 	``over'' or ``under'' the other, as in the last figure or the next one. 
But in Yang-Baxter theory, two lines
simply cross in the plane, as sketched here.}
\label{Crossing}
\end{figure}

Perhaps the most obvious question about the Yang-Baxter equation is ``why'' solutions of this highly overdetermined equation exist.    There is another area in which one finds something
a lot like the Yang-Baxter equations.   This is the theory of knots in three dimensions.    Knots are often described in terms of
projections to the plane; two knot projections describe the same knot if they can be related by certain moves that are known
as Reidemeister moves.  The most important Reidemeister move is shown in fig. \ref{Reidemeister}.
The resemblance to the Yang-Baxter equation is obvious, but there is also a conspicuous difference.
In knot theory, one strand passes ``over'' or ``under'' the other, while Yang-Baxter theory is a purely two-dimensional theory in which lines
simply cross, with no ``over'' or ``under,'' as sketched in fig. \ref{Crossing}.
Likewise, rough analogs of the other Reidemeister moves exist in Yang-Baxter theory (see figs. 4 and 5 of \cite{PW}), but they do not really have the same content as in knot theory
because
of the lack of a distinction between ``over'' and ``under.''

In this respect, knot theory has structure that is absent in  Yang-Baxter theory.   But there is also an important difference
in the opposite direction:  in Yang-Baxter theory, the spectral parameter is crucial, but it has no analog in knot theory.

Despite these differences, the analogy between the Yang-Baxter equation and the first Reidemeister move of knot theory is rather
conspicuous, so let us pursue this a little
bit.   The usual solutions of Yang-Baxter depend, as I have  said, on the choice of a simple Lie group\footnote{In the following
gauge theory construction, we take the compact form of $G$.  Also, we slightly simplify the exposition by assuming that $G$ is
connected and simply-connected.  This implies in particular that a $G$-bundle $E\to M$, where $M$ is a three-manifold,
is in fact trivial.  Hence a connection $A$ on $E$ can be understood as a Lie algebra valued 1-form, and $\CS(A)$ can be defined
by the naive formula (\ref{naive}).} $G$ and a representation $\rho$.   There
are knot invariants that depend on the same data.  To define them at least formally, let $M$ be a three-manifold, 
$E\to M$ a $G$-bundle, and $A$ a connection on $G$.
  Then one has the {\it Chern-Simons function}
  \be\label{naive}\CS(A)=\frac{1}{4\pi}\int_M\Tr\,\left(A\d A+\frac{2}{3}A\wedge A\wedge A\right),\ee
which was introduced in quantum field theory in \cite{Schwarz,Templeton}.   I have normalized $\CS(A)$
 so that, if $G=\mathrm{SU}(N)$ and
$\Tr$ is the trace in the $N$-dimenisonal representation, it is 
gauge-invariant mod $2\pi\Bbb Z$.   For any $G$, we define $\Tr$ as an invariant quadratic form on the Lie algebra of $G$  normalized
so that $\CS(A)$ is gauge-invariant predcisely mod $2\pi\Bbb Z$.
In quantum mechanics, the ``action''  $I$ must be well-defined mod $2\pi\Bbb Z$, so we can take
\be I=k\CS(A),~~~k\in\Bbb Z.\ee

  \begin{figure}
 \begin{center}
   \includegraphics[width=2.2in]{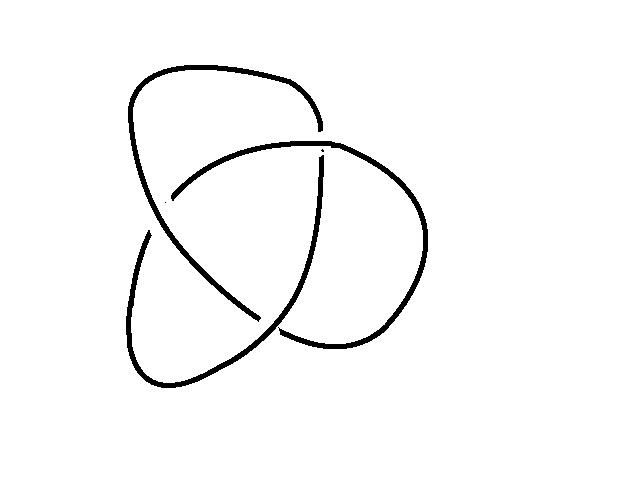}
 \end{center}
\caption{\small The trefoil knot, projected to the plane.}
\label{Knot}
\end{figure}
A quantum field theory with this action is a ``topological quantum field theory'' (modulo some subtleties involving
a framing anomaly), since there is no metric tensor in sight.    Let us just take the three-manifold $M$
to be $\Bbb R^3$, and let $K\subset \Bbb R^3$ be an embedded knot, such as the trefoil knot of fig. \ref{Knot}.
  We pick a representation $\rho$ of $G$, and let
\be W_\rho(K)=\Tr_\rho P\exp\left(\oint_KA\right)\ee
be the  Wilson loop operator (the trace of the holonomy) in the representation $\rho$.

The usual ``quantum knot invariants,'' of which the prototype is the Jones polynomial of a knot, can be defined via the
expectation value of the Wilson operator, $\left\langle W_\rho(K)\right\rangle =\left\langle \Tr_\rho P\exp(\oint_KA) \right\rangle$.   One can make knot
invariants this way, but from them  one cannot really extract the usual solutions of the Yang-Baxter equation
since one is missing the spectral parameter.    However (see \cite{SpecialCase}, eqn. (6.11)), in a sense from these knot invariants one can extract 
 a special case of the trigonometric solutions of the Yang-Baxter equation
in which the spectral parameter is taken to $\i\infty$.

How can we modify or generalize Chern-Simons gauge theory to include the spectral parameter?   A naive idea is to replace the finite-dimensional
gauge group $G$ with its loop group $\LG$.  We parametrize the loop by an angle $\theta$.   The loop group has ``evaluation'' representations that ``live''
at a particular value $\theta=\theta_0$ along the loop.\footnote{It is important here that by $\LG$ we mean the loop group, not its
central extension that is encountered in conformal field theory.  $\LG$, unlike its central extension, has homomorphisms to $G$
that map a loop $g(\theta)$ to its value $g(\theta_0)$ at some given $\theta_0$.  By composing this with an ordinary representation of $G$,
we get a representation of $\LG$ that informally ``lives'' at the point $\theta=\theta_0$.  These representations have  no close
 analog for the centrally extended loop group.}   We hope that this will be the spectral parameter label $\theta_0$ carried by a particle in the solution of the 
Yang-Baxter equation. 

Taking the gauge group to be a loop group means that the gauge field $A=\sum_i A_i(x) \d x^i$
now depends also on $\theta$ and so is $A=\sum_i A_i(x,\theta) \d x^i$.   Note that there is no $\d\theta$ term so this is not a full four-dimensional gauge field. 
The Chern-Simons action has a generalization to this situation:
\be\label{nthe} I=\frac{k}{4\pi}\int_{M\times S^1}\d\theta \,\Tr\left(A\d A+\frac{2}{3}A\wedge A\wedge A\right).\ee
This is perfectly gauge-invariant.  

\begin{figure}
 \begin{center}
   \includegraphics[width=2.9in]{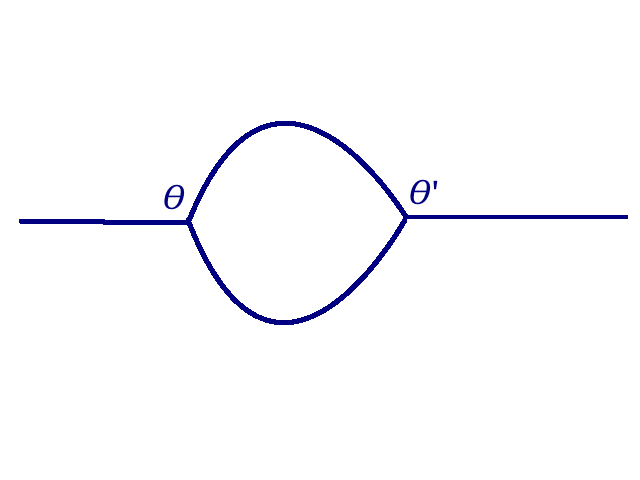}
 \end{center}
\caption{\small In a naive Chern-Simons theory with gauge group $\LG$, this loop will come with a factor  $\delta(\theta-\theta')^2=\delta(\theta-\theta')\delta(0)$ because of the two propagators connecting the same two vertices.}
\label{Loop}
\end{figure}

  What goes wrong is that because there is no  $\partial/\partial \theta$ in the action, the ``kinetic energy'' of $A$ is not elliptic
and the perturbative expansion is not well-behaved.    The propagator is
\be \left\langle A_i(\vec x,\theta) A_j(\vec x',\theta')\right\rangle=\frac{2\pi}{k}\frac{\epsilon_{ijk}(x-x')^k}{|\vec x-\vec x'|^2} \delta(\theta-\theta')\ee
with a delta function because of the missing $\partial/\partial\theta$. 
Because of the delta function, every loop diagram will acquire a factor  $\delta(0)$ (fig. \ref{Loop}).

Costello cured this problem via a very simple deformation.   Take our three-manifold to be $\Bbb R^3$, and
write $x,y,t$ for the three coordinates of $\Bbb R^3$,
so overall we have $x,y,t,$ and $\theta$.      Costello combined $t$ and $\theta$ into a complex variable
\be \label{zobb} z=\varepsilon t+\i \theta.\ee
Here $\varepsilon$ is a real parameter.  The theory will reduce to the bad case that I just described if $\varepsilon=0$.    As soon as $\varepsilon\not=0$,
its value does not matter, since it can be eliminated by rescaling $t$ or $\theta$, and one can set $\varepsilon=1$.  The purpose
of including  $\varepsilon$ in eqn. (\ref{zobb}) was simply to explain in what sense we will be making an infinitesimal deformation
away from the ill-defined Chern-Simons theory of the loop group. 

One replaces $\d\theta$ (or $(\i k/4\pi)\d\theta$) in the naive theory (\ref{nthe}) with $\d z$ (or $\d z/\hbar$, where $\hbar$
is a parameter\footnote{One should think of $\hbar$ as a formal complex parameter, since there is in general
no natural reality condition that it might obey.  We return to this point at the end of these notes.}  that plays the role of $1/k$ in the usual Chern-Simons theory) and one now regards $A$ as a partial connection on
$\R^3\times S^1$ that is missing a $\d z$ term (rather than  missing $\d\theta$, as before).    The action is now
\be\label{ena} I=\frac{1}{\hbar}\int_{\R^3\times S^1}\d z\,\Tr\,\left(A\d A+\frac{2}{3}A\wedge A\wedge A\right). \ee

We have lost the three-dimensional symmetry of standard Chern-Simons theory, because of splitting away one of the three coordinates of $\R^3$
and combining it with $\theta$.   We still have two-dimensional diffeomorphism symmetry.   However, as we discussed in comparing Yang-Baxter theory to knot theory,
Yang-Baxter theory does not have three-dimensional symmetry, but only two-dimensional symmetry.    Modifying standard Chern-Simons theory
in this fashion turns out to be just right to give Yang-Baxter theory rather than knot theory:  the three-dimensional diffeomorphism invariance
is reduced to two-dimensional diffeomorphism invariance, but on the other hand, now there is a complex variable $z$ that will turn out to be the (complexified) spectral parameter.

I have described the action so far on $\R^2\times \C^*$ where $\C^*=\R\times S^1$ (parametrized by $z=t+\i \theta$) is endowed with the complex 1-form $\d z$.
  The classical action makes sense more generally on\footnote{\label{framing} 
  Quantum mechanically, one runs into
  an analog of the framing anomaly of Chern-Simons theory, such that $\Sigma$ has to be framed (its
  tangent bundle has to be trivialized) in order to define the quantum theory.  This is very restrictive, since
  a  compact two-manifold that can be framed is topologically  a two-torus $T^2$. The
  example $\Sigma=T^2$ with a periodic array of Wilson lines is used in \cite{Costello} in
  constructing integrable lattice models with periodic
  boundary conditions.  There is also an analog in this theory of the usual framing anomaly for
  knots in Chern-Simons theory; this anomaly markedly restricts the class of Wilson operators that can
  be considered.  For example, a Wilson operator supported on a simple closed loop in $\Sigma=\R^2$,
  at a point in $C$, is anomalous. There is no problem for the ``straight'' Wilson lines that are used in
  constructing integrable lattice models.}
     $\Sigma\times C$, where $\Sigma$ is any smooth (oriented) two-manifold and $C$ is a complex
Riemann surface endowed with a holomorphic 1-form $\omega$:
\be\label{gena} I=\frac{1}{\hbar}\int_{\Sigma\times C}\omega \wedge \Tr\,\left(A\wedge \d A+\frac{2}{3}A\wedge A\wedge A\right). \ee
 It turns out, however, that to get a quantum theory, one wants $\omega$ to have no zeroes.    Intuitively this is because a zero of $\omega$
is equivalent to a point at which $\hbar\to\infty$.  (More technically, if  $\omega$ has a zero, the kinetic
energy of the theory is not elliptic and one runs into a difficulty in quantization
somewhat similar to the problem with the 
naive theory of eqn. (\ref{nthe}).  One can resolve the problem by modifying the theory near a zero of $\omega$,
but then two-dimensional symmetry is lost.)   By constrast, there is no problem 
with poles of $\omega$, where effectively
$\hbar\to 0$.

So $C$ has to be a complex Riemann surface that has a differential $\omega$ with possible poles, but with no zeroes.  
The only three options are $\C$, $\C/\Z\cong \C^*$, and $\C/(\Z+\tau\Z)$, which  (with $\tau$ a complex number of positive imaginary part) is a Riemann surface of genus 1.
  It turns out that these three cases correspond to the three traditional classes of solutions of the Yang-Baxter equation -- rational,
trigonometric, and elliptic.  

The first point is that this theory has a sensible propagator and a sensible perturbation expansion.    The basic reason for a sensible propagator is that on $\R\times \R$ or $\R\times S^1$
parametrized by $t$ and $\theta$, the operator $\partial/\partial t$ that appeared in the naive action (\ref{nthe})
 is not elliptic, but the operator $\partial/\partial\overline z$
that appears in the deformed version is elliptic.   After a suitable gauge-fixing, the propagator (for the rational model, i.e. on $\R^2\times C$ with $C=\C\cong \R^2$) is
\be \langle A_i(x,y,z)A_j(x',y',z')\rangle=\hbar \varepsilon_{ijk z}g^{kl}\frac{\partial}{\partial x^l}\left(\frac{1}{(x-x')^2+(z-z')^2 +|z-z'|^2}\right),\ee
where  $\varepsilon_{ijkm}$ is the four-dimensional antisymmetric tensor (but in the formula we set $m=z$, so that $i,j,k$ take the values $x,y,\bar z$),
 and the metric on $\R^4=\R^2\times C$ is $\d x^2+\d y^2+|\d z|^2$.

With this propagator, the perturbative expansion is well-defined, as shown in \cite{Costello}.    This is a tricky point.  The theory is actually unrenormalizable by power counting,
so on that basis, one would not expect a well-behaved quantum theory.    However, it has no possible counterterms, because all local gauge-invariant operators
vanish by the classical equations of motion.   Anyway, using a fairly elaborate algebraic machinery of BV quantization, it is shown in \cite{Costello} that the theory has a well-defined
perturbation expansion.

Now we consider Wilson operators, that is holonomy operators
\be \Tr_\rho P\exp\oint_\ell A\ee
where $\ell$ is a loop in $\Sigma\times C$.  As before, $\Sigma$ is a topological two-manifold, and $C$ is a complex Riemann surface with the differential $\omega=\d z$. 
But we have only  a partial connection
\be A=A_x \d x + A_y \d y +A_{\bar z}d\bar z\ee
so we would not know how to do any parallel transport in the $z$ direction.    (We cannot interpret $A$ as a gauge field with $A_z=0$ because this condition
would not be gauge-invariant, and quantizing the theory requires gauge-invariance.  We have to interpret it as a theory with $A_z$ undefined, so we cannot do parallel transport
in the $z$ direction.)   This means that we must take $\ell$ to be a loop that lies in $\Sigma$, at a particular value of $z$. 

But this is what we wanted to explain Yang-Baxter theory.  It means that $\ell$ is labeled by some constant
value $z=z_0$ of the (complexified) spectral parameter $z$.

  \begin{figure}
 \begin{center}
   \includegraphics[width=2.9in]{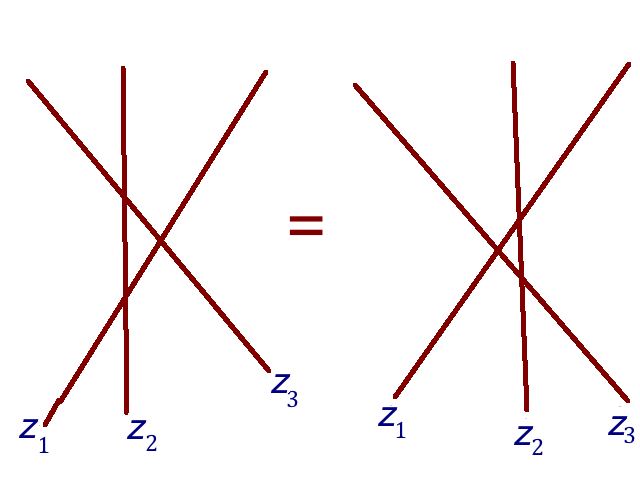}
 \end{center}
\caption{\small Three lines crossing in the plane, now labeled by the value of 
a complexified spectral parameter $z$.  The equivalence of the two pictures is the Yang-Baxter equation.}
\label{YB2}
\end{figure}

Now let us consider some lines  that meet in $\Sigma$ in the familiar configuration associated to the Yang-Baxter equation (fig. \ref{YB2}).
Two-dimensional diffeomorphism invariance means that we are free to move the lines around as 
long as we do not change the topology of the configuration.
  But assuming that $ z_1, $ $z_2$, and $z_3$ are all distinct, it is manifest that there is no 
  discontinuity when we move the
middle line from left to right even when we do cross between the two pictures, because the Feynman diagrams
in four dimensions have no singularity when this occurs.   Thus two configurations of
Wilson operators that differ by what we might call a Yang-Baxter move are equivalent.

\begin{figure}
 \begin{center}
   \includegraphics[width=2.9in]{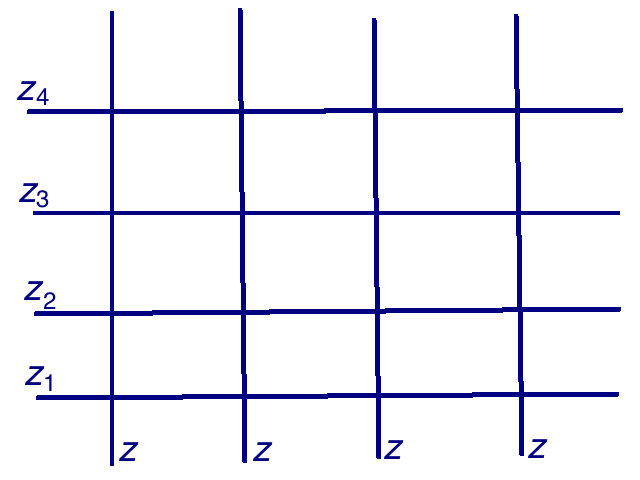}
 \end{center}
\caption{\small The configuration of Wilson lines associated to the integrable lattice models.
The vertical and horizontal lines are labeled by points in $C$ and by representations of $G$ (the representations
are not indicated in the figure). }
\label{Lattice2}
\end{figure}

Likewise, in the configuration associated to integrable lattice models
(fig. \ref{Lattice2}),
we can move the horizontal lines up and down at will.

But why is there as elementary a picture as in the usual integrable lattice models, where one can 
evaluate the path integral by labeling
each line by a basis element of the representation $\rho$ and each crossing by a local factor $R_{ij,kl}(z)$?
This is a little tricky and depends on picking the right boundary conditions so that the only classical solution of
the equations of motion
is the trivial solution $A=0$, and it has no automorphisms (unbroken gauge symmetries).
There is a way to achieve this for each of the three choices of $C$,
corresponding to rational, trigonometric, and elliptic solutions of the Yang-Baxter equation.

For simplicity I will here only explain the rational case, 
in which the Riemann surface is $C=\C$, the complex $z$-plane, with 1-form $\d z$.   
We require that the gauge field $A$ on $\Sigma\times \C$ 
goes to 0 at infinity in the $\C$ direction, and likewise in quantizing we divide only by gauge 
transformations that approach 1 at infinity along $\C$. 
With these conditions, it is indeed true that $A=0$ is the only classical solution, and that
it has no unbroken gauge symmetries.    However, $G$ still exists as a group of global symmetries (corresponding
to gauge transformations that are constant at infinity along $C$, rather than equaling 1).
That will explain, from this point of view, why the rational solutions of the Yang-Baxter equation, which
arise from this construction, are $G$-invariant.\footnote{For the other choices of $C$,
boundary conditions that allow a unique classical solution that is free of unbroken gauge symmetries
do exist (for any $G$ if $C=\C^*$ and for $G=PSU(N)$ if $C$ is a curve of genus 1), but lack the full global $G$ symmetry.
This explains from the present point of view why the trigonometric and elliptic solutions of Yang-Baxter
are not $G$-invariant, and why elliptic solutions only exist for the special linear group.}

When  we expand around the trivial solution $A=0$, 
the absence of deformations or automorphisms of this solution makes
 the perturbative expansion  straightforward.     
Moreover, perturbation theory gives a simple answer because the theory is infrared-trivial, which is the flip
side of the fact that it is unrenormalizable by power-counting.   That 
means that effects at ``long distances'' in the topological space are negligible.

I put the phrase ``long distances'' in quotes because two-dimensional 
diffeomorphism invariance means that there is no natural notion of distance on
the topological two-manifold 
$\Sigma$.   A metric on $\Sigma\times \C$ entered only when we fixed the gauge to pick
a propagator.  Recall that we used the metric $\d x^2+\d y^2 +|\d z|^2$.    We could equally well scale up the metric along $\Sigma$ by any factor
and use instead $e^{B}(\d x^2+\d y^2)+|\d z|^2$ for very large $B$.  

  \begin{figure}
 \begin{center}
   \includegraphics[width=2.9in]{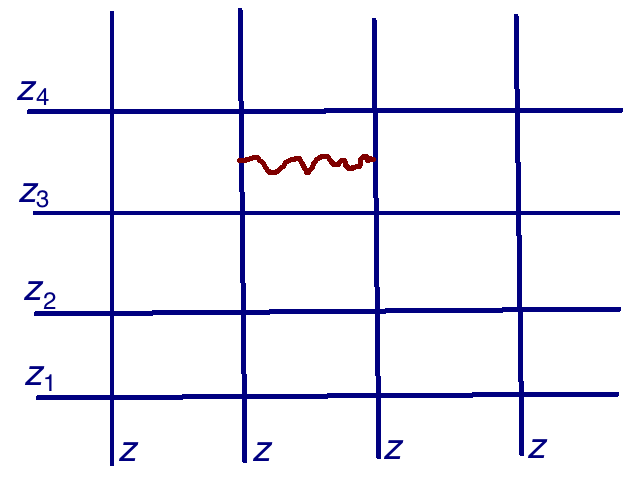}
 \end{center}
\caption{\small Because the deformed Chern-Simons theory is defined with no choice of metric on $\Sigma$,
we can always  use a gauge such that nonintersecting Wilson lines are very far apart.  Gauge boson exchanges between nonintersecting Wilson lines, as depicted here, vanish in this limit. }
\label{Lattice3}
\end{figure}

That means that when one looks at the picture of fig. \ref{Lattice2} associated to the integrable lattice systems,
one can consider the vertical lines and likewise the horizontal lines to be very far apart (compared to $z-z_i$ or $z_i-z_j$). 
 In such a situation, in an infrared-free theory, effects that involve a gauge boson exchange between two nonintersecting lines, as in fig. \ref{Lattice3},
are negligible.

  \begin{figure}
 \begin{center}
   \includegraphics[width=2.9in]{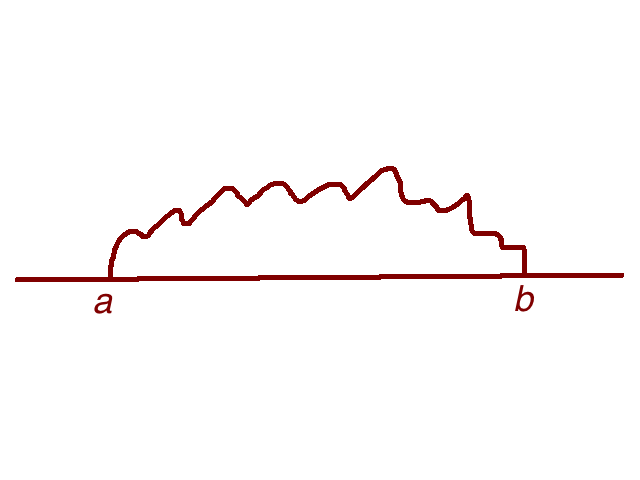}
 \end{center}
\caption{\small  Exchange of a gluon from one Wilson line to itself would correspond to mass renormalization
(of an external probe charge) in conventional quantum field theory.  In the framework that leads
to the integrable lattice models, the symmetries do not allow any nontrivial effect analogous to mass
renormalization. }
\label{MassRenormalization}
\end{figure}

One should worry about gauge boson exchange from one line to itself, as in fig. \ref{MassRenormalization}, 
 because in this case the distance $|a-b|$ need not be large.    Such effects correspond roughly
to ``mass renormalization'' in standard quantum field theory.    In the present problem, in the case
of a straight Wilson line such as those of fig. \ref{Lattice2}, the symmetries do not allow any interesting effect analogous to mass renormalization.   (For more general Wilson lines, the  ``mass renormalization'' diagram
leads to a framing anomaly that was mentioned in footnote \ref{framing}.)

  \begin{figure}
 \begin{center}
   \includegraphics[width=2.9in]{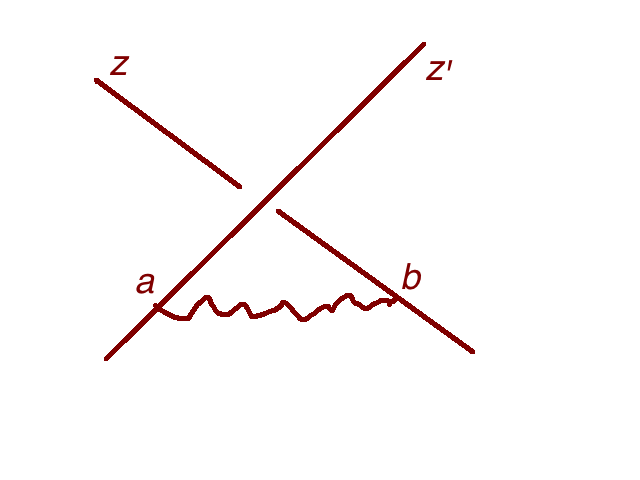}
 \end{center}
\caption{\small The lowest order diagram with gauge boson exchange between two lines that cross in
$\Sigma$.  The two lines have   spectral
parameters $z,z'$.  To evaluate the contribution of this diagram, we have to integrate over $a$ and $b$.
The integral converges and is dominated by the region $|a|,|b|\lesssim |z-z'|$. }
\label{FirstCorrection}
\end{figure}
What about gauge boson exchange between two Wilson lines with distinct spectral parameters $z,z'$ that do cross in $\Sigma$?  The
lowest order example with exchange of a single gauge boson is depicted in fig. \ref{FirstCorrection}.   To evaluate the contribution of this
diagram, we have to integrate over the points $a,b$ at which the gauge boson is attached to the two
Wilson lines.  The resulting integral converges and is dominated by the region $|a|,|b|\lesssim |z-z'|$.
What it converges to will be discussed shortly.

  \begin{figure}
 \begin{center}
   \includegraphics[width=4.2in]{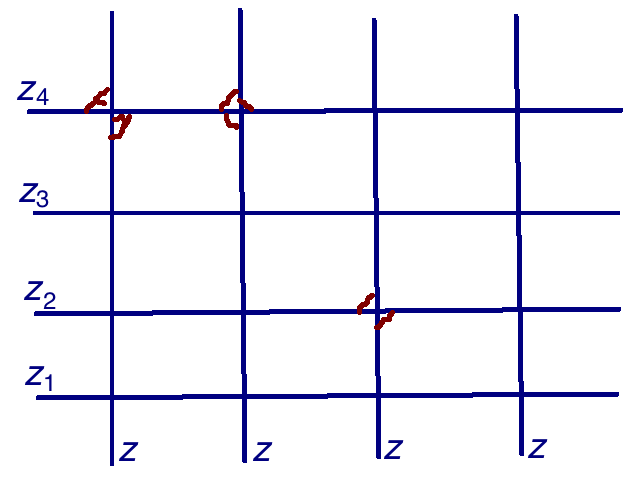}
 \end{center}
\caption{\small Evaluating in perturbation theory the configuration of Wilson operators associated to the
integrable lattice models, we run into very complicated Feynman diagrams.  But after scaling up the metric
of $\Sigma$, we can assume that these diagrams are built from decoupled diagrams each of which is
associated to just one crossing point.}
\label{Lattice4}
\end{figure}

Now when we study a general configuration such as the one related to the integrable lattice models we can draw very complicated diagrams, as in fig. \ref{Lattice4},
but the complications are all localized near one crossing point or another.
The diagrams localized near one crossing point simply build up an $R$-matrix associated to that crossing, and the discussion makes
it obvious that the $R$-matrix obeys the Yang-Baxter relation of fig. \ref{YB2}.

Moreover, this makes it clear that the path integral in the presence of the configuration of Wilson operators 
(fig. \ref{Lattice2})
associated to the integrable
lattice models can be evaluated by the standard rules: 
label each vertical or horizontal line segment by a basis vector of the representation $\rho$ and
include the appropriate $R$-matrix element at each crossing; then sum over all such labelings.

But why is the $R$-matrix obtained this way the standard rational solution of the Yang-Baxter equation?   (And
similarly, with a different choice of $C$, why would we get the standard trigonometric and elliptic solutions
of Yang-Baxter?)
 In his paper \cite{Costello}, Costello explicitly evaluates the lowest order correction in
\be\label{zelf} R=1+\hbar r +{\mathcal{O}}(\hbar^2)\ee
 from the lowest order non-trivial diagram of fig. \ref{FirstCorrection}, and 
 and gets the standard answer
 \be r= \frac{\sum_a t_a t'_a}{z-z'} \ee
 (where $t_a$, $t_a'$, $a=1,\dots,{\mathrm{dim}}\,G$ are the generators of the Lie algebra of $G$ acting in the two representations).    Once the first
 order deformation is known, the full answer for the $R$-matrix, up to a change of variables,
  follows from general arguments (see p. 814 in \cite{Drinfeld} and p. 418 in \cite{Guide}).
 
I conclude with a few final comments.
 Costello's theorem is purely about perturbation theory, but his theorem shows that, in this particular theory (and rather exceptionally),
  perturbation theory converges.    As a physicist, one would want to
 give an {\it a priori}  ``nonperturbative definition'' of the theory, which would have the claimed perturbative expansion.   At first sight, there are some difficulties in doing this.  The action of conventional Chern-Simons
 theory is gauge-invariant mod $2\pi\Z$ if the coupling parameter $k$ is an integer, as one wishes
 for a quantum system that will be well-defined beyond perturbation theory.  But there is no choice of
 the analogous parameter $\hbar$ in eqn. (\ref{ena}) or (\ref{gena}) that makes the action gauge-invariant
 mod $2\pi \Z$.  Also, the path integral of conventional Chern-Simons gauge theory is an oscillatory path
 integral, like that of any unitary quantum system, because the  action is real.\footnote{With $\CS(A)$ defined
 as in eqn. (\ref{naive}), the argument of the path integral is $\exp(\i k\CS(A))$.}  The real
 and imaginary parts of the generalized action in eqn. (\ref{ena}) or (\ref{gena}) are unbounded above and below,
 so naively the path integral is exponentially divergent, no matter what  we assume for $\hbar$.
 
 Because of such considerations, to get a convergent, nonperturbative path integral in this theory, one 
 will have to consider an analytically continued
 version of the theory, in a sense that for conventional Chern-Simons theory was described in \cite{FiveBranes}.
 This analytic continuation is achieved by complexifying all variables and then constructing a different integration
 cycle on which the path integral converges.
 For ordinary Chern-Simons theory, such analytic continuation is optional;  the theory at integer $k$ is perfectly
 well-defined without any analytic continuation.  But in the four-dimensional
 theory that is under discussion here, only an analytically-continued
 version of the path integral will make sense beyond perturbation theory.  To construct this analytic
 continuation, one can consider the D4-NS5 system of string theory, and proceed via the same arguments that were used
 in \cite{FiveBranes} to relate ordinary Chern-Simons theory and the Jones polynomial to the D3-NS5 system.
 
 An interesting goal for the future is to somehow link the story
  that has been reviewed here to the work of Nekrasov and Shatashvili,
 who developed 
 a seemingly quite different relationship between integrable quantum spin systems and supersymmetric gauge
 theory \cite{NS,NS2}.
 
 Another natural question is whether models such as the chiral Potts model (see for example \cite{YM,BPA,BaS,BBP})
 can be placed in the gauge theory framework. The chiral Potts model is an integrable lattice model
 in which the spectral parameter takes values in a curve of genus greater than 1.  Perhaps surprisingly (given the genus of its spectral curve),
 the chiral Potts
  model can be related to a trigonometric solution
 of Yang-Baxter for the group $SU(2)$ (or $SL_2$), and this viewpoint has also been used to develop an analog for $SU(N)$
 \cite{DJMM,BKMS}.
 But such models are not yet understood in the framework described in the present lecture.

Research supported in part by NSF Grant PHY-1606531.   I thank K. Costello, J. Perk, and M. Yamazaki for useful comments.

\bibliographystyle{unsrt}

\end{document}